\documentstyle[prl,aps, multicol, graphicx]{revtex}
\begin{document}
\draft
\widetext

\title{Anisotropies of the lower and upper critical fields
in MgB$_2$ single crystals}

\author{L.Lyard$^1$, P.Szab\' o$^{1,2}$, T.Klein$^{1,3}$, J.Marcus$^1$,
C.Marcenat$^4$, B.W.Kang$^5$, K.H.Kim$^5$, H.S.Lee$^5$ and S.I.Lee$^5$}
\address{$^{1}$ Laboratoire d'Etudes des Propri\'et\'es Electroniques des
Solides, CNRS, B.P.166, 38042 Grenoble Cedex 9, France}
\address{$^2$  Centre of  Low Temperature  Physics IEP  SAS \& FS
UPJ\v S, Watsonova 47, 043 53 Ko\v{s}ice, Slovakia   }
\address{$^{3}$ Institut Universitaire de France and Universit\'e Joseph
Fourier, B.P.53, 38041 Grenoble Cedex 9, France}
\address{$^{4}$ Commissariat \`a l'Energie Atomique - Grenoble,
D\'epartement de Recherche Fondamentale sur la Mati\`ere Condens\'ee,
SPSMS, 17 rue des Martyrs, 38054 Grenoble Cedex 9, France}
\address{$^5$ NVCRICS and department of Physics, Pohang University of Science and Technology, Pohang 790-784 Republic of korea}
\date{\today}

\maketitle

\begin{abstract}
The  temperature  dependence  of  the  London  penetration  depth
($\lambda$) and  coherence length ($\xi$)  has been deduced  from
Hall  probe magnetization  measurements in  high quality  MgB$_2$
single crystals  in the two main  crystallographic directions. We
show that,  in contrast to  conventional superconductors, MgB$_2$
is   characterized   by   two   different  anisotropy  parameters
($\Gamma_\lambda  =  \lambda_c/\lambda_{ab}$  and  $\Gamma_\xi  =
\xi_{ab}/\xi_c$)  which strongly  differ at  low temperature  and
merge at $T_c $. These results are in very good agreement with
recent calculations in weakly coupled two bands suprerconductors (Phys. Rev. B, 66, 020509(R) (2002).
\end{abstract}


\begin{multicols}{2}
\narrowtext

The recent  discovery of superconductivity  in Magnesium Diboride
\cite{nagamatsu} has been the starting point of a large number of
theoretical    and   experimental    investigations.   The coexistence of two
superconducting   energy   gaps   with   different   anisotropies
\cite{szabo} is  expected to  give rise  to very  peculiar
physical  properties in this system \cite{liu,kogan,dahm}.  Among  those,  the  strong
temperature dependence  of the anisotropy  of the upper  critical
field $\Gamma_{ H_{c2}} = H_{c2 \|ab}/H_{c2 \|c}$ is now well
established \cite{lyard,angst} ($H_{c2 \|ab}$ and $H_{c2 \|c}$
being the upper critical fields for magnetic fields parallel to
the two main crystallograpic directions, $ab-$planes and $c-$direction,
respectively).\

"Classical"  superconductors can  be characterized  by one single
anisotropy    parameter    :    $\Gamma    =   \xi_{ab}/\xi_c   =
\lambda_c/\lambda_{ab}$  (where  $\xi$   and  $\lambda$  are  the
superconducting  coherence  and   penetration  depths  in  signed
crystallographic  orientations)  whatever  the  anisotropy of the
superconducting   gap.  However,   it  has   been  suggested  that
$\xi_{ab}/\xi_c$      could     differ      considerably     from
$\lambda_c/\lambda_{ab}$ at low temperature in MgB$_2$ due to the
presence of two superconducting energy gaps with  different anisotropies \cite{kogan}.
Even though recent
neutron      scattering     data      have     confirmed     that
$\lambda_c/\lambda_{ab} \sim 1$ at low temperature is indeed very
different  from  $\xi_{ab}/\xi_c   \sim  5$  \cite{neutron},  the
temperature       dependence      of       $\Gamma_\lambda      =
\lambda_c/\lambda_{ab}$  still  had  to  be  determined.  In this
letter,  we  confirm  that  $\Gamma_\xi = \xi_{ab}/\xi_c$ ($=\Gamma_{H_{c2}}$)  strongly  differs  from $\Gamma_\lambda$ at  low temperature and show  for the first time
that $\Gamma_\xi$ and $\Gamma_\lambda$ present different temperature
dependencies : whereas $\Gamma_\xi$ decreases from $\sim 5$ to $\sim 2$
(close to $T_c$), $\Gamma_\lambda$ increases with temperature tending
from $\sim 1.4$ towards $\Gamma_\xi$ for $T \rightarrow T_c$.\

Magnetization  measurements have  been performed  on high quality
single  crystal \cite{sungik}  ($T_c \approx 36.5$ K) showing flat  surfaces of  typical
dimensions   :   $100*100*25   \mu$m$^3$   using   a  Hall  probe
magnetometer.  The main  surface  (i.e.  the $ab-$planes)  of the
sample has been placed either parallel to or perpendicular to the
surface of the Hall probes  in order to measure the magnetization
for $  H \| c$  and $H \|  ab$ respectively (see  sketches in the
inset of Fig.1a).  The alignement of the external  field with the
main crystallographic axis  has been obtained  by slightly rotating  the
ensemble in order to get  the maximum and mimimum $H_{c2}$ values
for $H \| ab$ and $H  \| c$ respectively. A typical magnetization
loop is  presented in the  inset of Fig1b  for $H \|  c$ and $T =
20$ K. As shown the  irreversible part of the magnetization loops
is rather small (corresponding to  critical current values on the
order of $10^3$  A/cm$^2$ at low T and low  H) and the reversible
magnetization could thus easily be obtained assuming that
$M_{rev} = (M_{up} + M_{down})/2$ where $M_{up}$ and $M_{down}$
are the magnetization for increasing and decreasing magnetic fields
respectively (Bean critical state model). Typical curves are displayed in Fig.1 for $T > 26K$.
For lower temperatures, the upper critical field values for
$H \| ab$ became larger than our maximum field ($3$ T) but the
lower critical field could still be deduced easily in both directions
down to the lowest temperatures ($5$ K).\

For  $H  \|  c$  i.e.  perpendicular  to  the platelet, important
demagnetizing effect  come into play  and the magnetic  field has
been  rescaled to  : $H  = H_0  - N_cM_{rev}$  where $H_0$ is the
external field  and $N_c$ the demagnetizing  factor. $N_c$ can be
estimated assuming  that the sample is  an ellipsoid of thickness
$t \sim  25 \mu$m and  width $w \sim  100 \mu$m to  $N_c = 1 -\pi
t/2w \approx 0.6$. This value has been confirmed by the fact
that, after  correction  for  this  demagnetization  effect, the
magnetization curves present a nearly vertical slope for
$H = H_{c1}$. As pointed out by Zeldov {\it et al.} \cite{zeldov}
this value might however be overestimated by a factor on the order of
$\sqrt{w/t}$ due to the presence of geometrical barriers but it is
important to note that, even though the absolute values of
$H_{c1 \| c}$ - and $H_{c1  \| ab}$ - are directly related to the
choice of the demagnetizing factor, their temperature dependence
is absolutely not affected by this choice. \

\begin{figure}
\includegraphics [width=7cm]{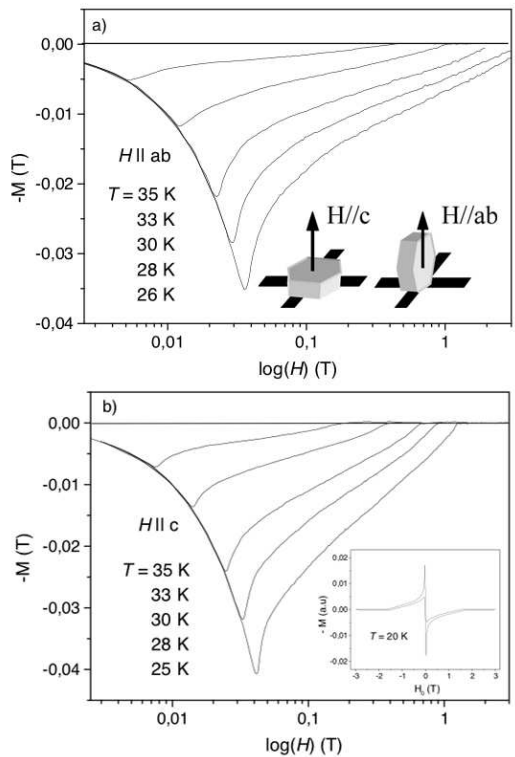}
\caption{Magnetic field  dependence of  the
reversible magnetization for $H \|c$ (a) and $H \|ab$ (b) at various temperatures. In the inset : (a): schematical drawings of the geometry used for both directions, and (b): complete magnetization loop at $T = 20$ K for $H \|c$. }
\end{figure}

The value of $N_{ab}$ has then been set to $N_{ab} \approx 1 - N_c/2$.
In this direction, the effect of Bean Livingston barriers can also
lead to a significant overestimation of $H_{c1}$ (up to a factor
$\kappa/ln(\kappa)$ where $\kappa$ is the Ginzburg -Landau parameter, the penetration field is then equal to the thermodynamic field instead of the lower critical field).
Those effects play for instance  a significant role in high $T_c$
cuprates due to their very high $\kappa$ values \cite{konczykowski}.
They result from the competiting effect between an attracting image
force to the surface and a repulsive one  arising from the interaction  between the vortices
and the shielding currents. This  latter force is proportional to
the  magnetization and  thus vanishes  as $M  \rightarrow 0$  for
descending  magnetic  fields. When
the  field further  decreases  nothing  prevents the  vortices to
leave the  sample and $M$ remains  close to zero. This  effect is
highly sensitive to  the quality of the surface  and we did never
observe this characteristic behaviour in our magnetization loops.
On the contary the irreversible part of the magnetization remained
small in the entire temperature range (being for instance on the order of $15 \%$
at $30$ K) and we thus assumed that $M_{rev}$ is equal to the average magnetization (i.e.  that pinning mainly arises from bulk defects).\

The corresponding $M_{rev}$ versus $H$ curves are displayed in Fig.1.
Note that the y-axis has been slightly rescaled in order to get a $-1$
slope in the Meissner state due to the fact that the sample did not
completely recover the surface of the Hall probe (especially for $H \| ab$).
Assuming that the lower critical magnetic field is equal to the
penetration field (i.e. neglecting durface pinning effects, see discussion above), $H_{c1}$ could be easily  determined from  the well
defined minima in the $M_{rev} (H)$ curves (see Fig.1).
The upper critical magnetic field $H_{c2}$ has been defined as the onset
of the diamagnetic response at $M_{rev}(H_{c2})= 0$.\

\begin{figure}
 \includegraphics [width=7cm]{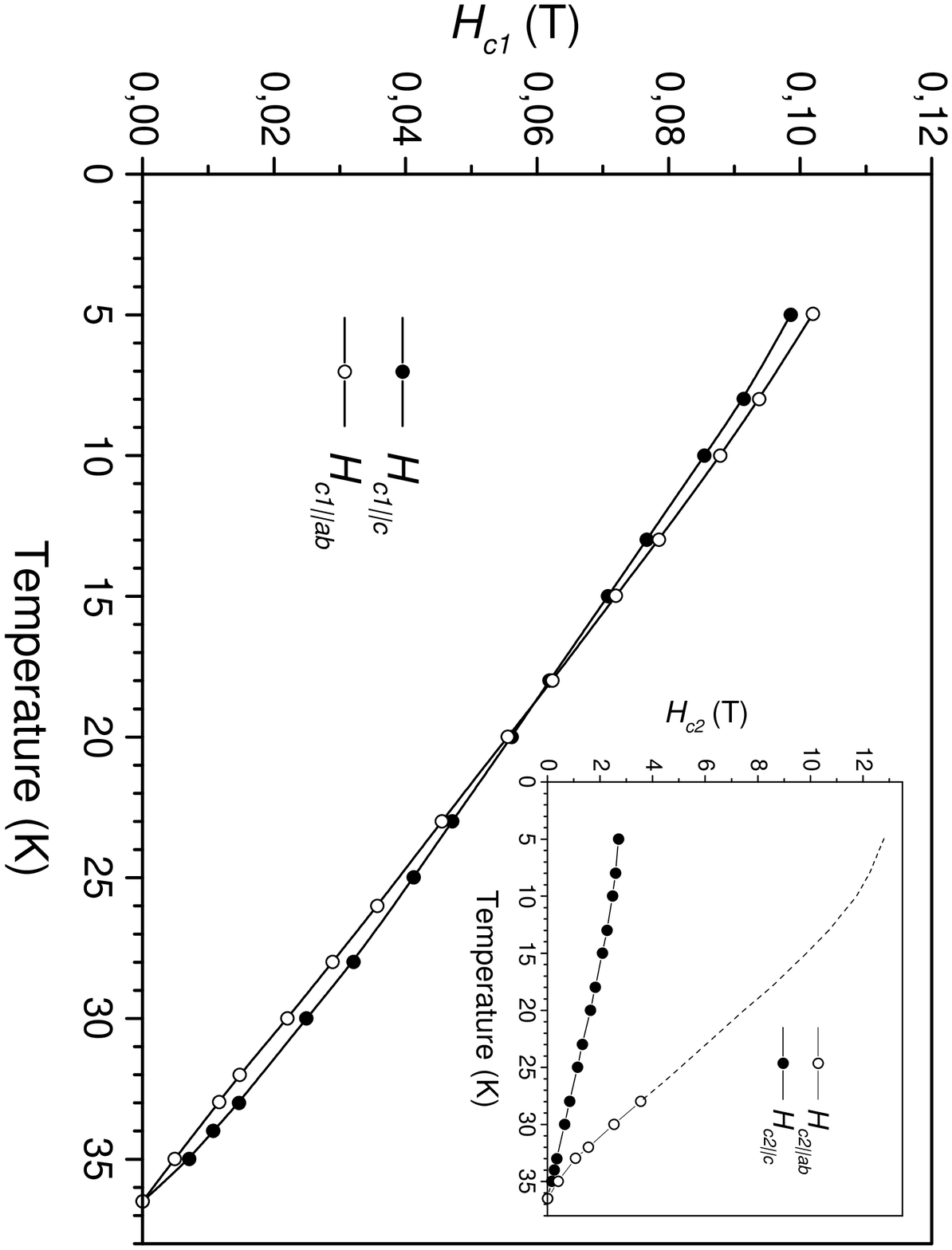}
 \caption{Temperature dependence of the lower critical field for $H \| c$ and $H \| ab$. In the inset : Temperature dependence of the upper critical field for the same directions. The dotted line is an extrapolation below $26$ K deduced from our previous magnetotransport measurements [5]}\end{figure}

The temperature dependencies of the corresponding critical fields
are  displayed   in  Fig.2.  The  inset   shows  the  temperature
dependence of the upper critical magnetic  field for $H \| c$ (solid circles) and
$H \|  ab$ (open circles) deduced  from our  magnetization measurements.  Those values have
been extrapolated down to low  temperature for $H \|ab$ using our
previous high  field magneto-transport data  performed on a  sample
from the  same batch \cite{lyard} (with  slightly higher $H_{c2}$
values,  dotted  line  in  the  inset  of  Fig.2).  As previously
observed   \cite{lyard,angst},  the   temperature  dependence  of
$H_{c2}$ is  almost linear for $H  \| c$ and presents  a positive
curvature  for  $H  \|  ab$.  As  shown,  the  situation  is very
different for the lower critical field which is almost linear for
$H \| ab$ and reveals a {\it  negative} curvature for $H \| c$ at
high temperatures. This unusual negative curvature of $H_{c1\| c}$
close to $T_c$ has been previously observed in thin films and
polycrystalline samples \cite{buzea} and can be explained by the
two-band Ginzburg-Landau theory \cite{askerzade}
(the isotropic Ginzburg-Landau theory only allows for a linear
temperature dependence of $H_{c1}$ close to $T_c$ \cite{abrikosov}).
As shown below, this difference in  curvatures will lead to an increase
of $\Gamma_\lambda$ for increasing temperature. \

\begin{figure}
 \includegraphics [width=7cm]{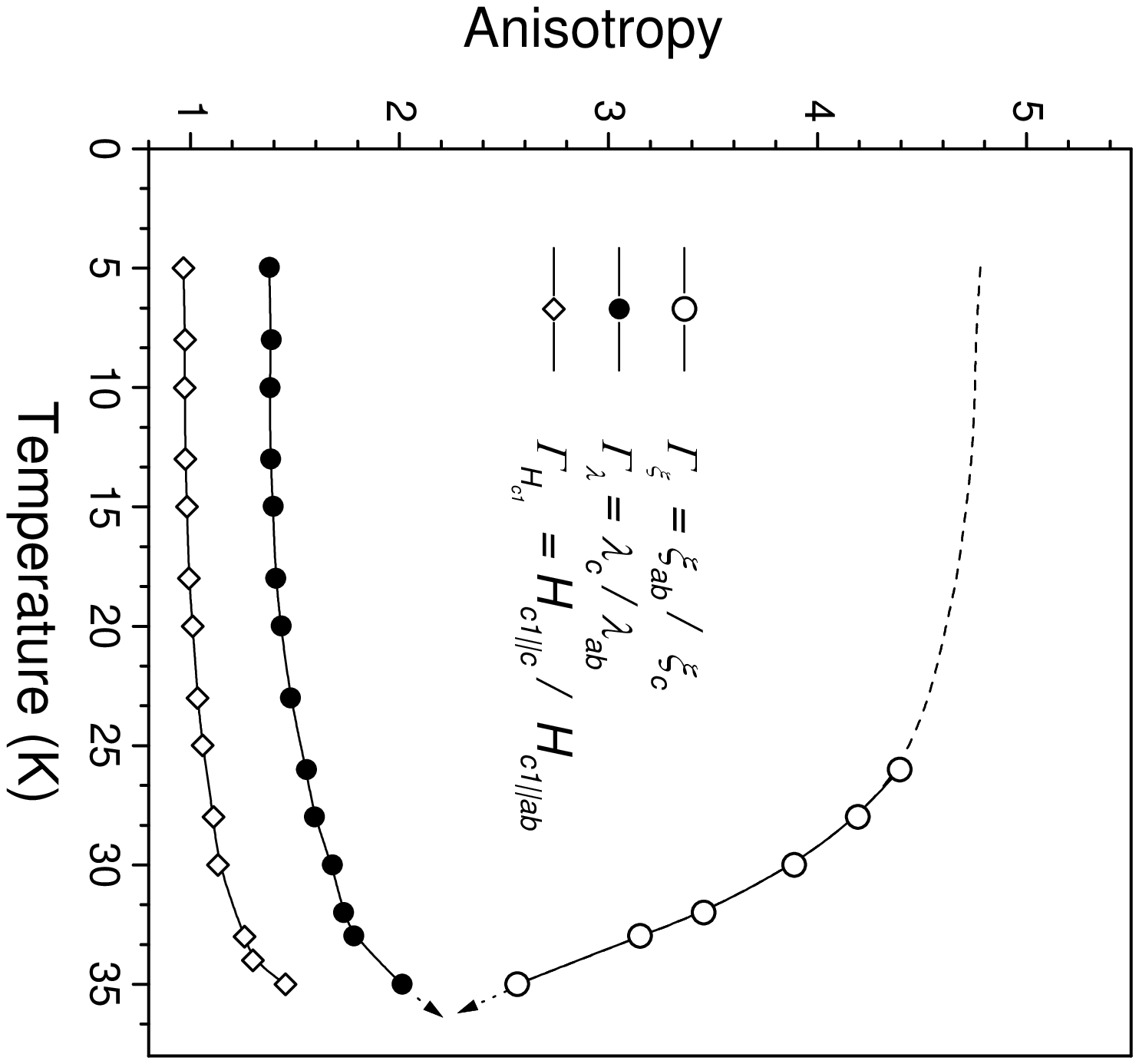}
\vskip -3cm
 \caption{Temperature dependence of the anisotropy parameters $\Gamma_\xi$, $\Gamma_\lambda$ and $\Gamma_{H_{c1}}$ (see text for details). The dotted line is an extrapolation of $\Gamma_\xi$ at low temperature deduced from our previous magnetotransport measurements [5].}\end{figure}

The penetration depth has been deduced from $H_{c1}$ through :
\begin{equation}
H_{c1} = \Phi_0/(2\pi \lambda_{ab}^2). (\Gamma_H/\Gamma_\lambda) .(ln(\tilde{\kappa}) + 0.5)
\end{equation}
with   $\tilde{\kappa}  =   \lambda_{ab}/\xi_{ab}$,  $\Gamma_H  =
\Gamma_\lambda$  for   $H  \|  c$   and  $\tilde{\kappa}  \approx
\sqrt{\lambda_{ab}\lambda_c/\xi_{ab}\xi_c}$,   $\Gamma_H =  1$ for
$H \| ab$. The values  of the coherence lengths $\xi_c(0) \approx
5$ nm  and  $\xi_{ab} (0) \approx 10$ nm  have been determined
from $H_{c2}(T)$ and we hence got $\lambda_c(0) \approx 80$ nm
and $\lambda_{ab}(0)  \approx 60$ nm  in good agreement  with the
values deduced from the slope of the magnetization curve :
$1/\lambda^2 = [8\pi/\Phi_0]d(\mu_0M_{rev})/dln(H)$.
Note that this slope rapidly decreases with field up to some
characteristic field $H^*(T)$ being on the order of $0.3$ T at low
temperature. This field can probably be associated with the closing
of the  small gap, $H^*  \sim H_{c2}^\pi$ in  good agreement with
our previous point-contact spectroscopy experiment \cite{samuely} and with
estimation of Bouquet {\it et al.} \cite{bouquet} from specific heat
measurements and Eskildsen {\it et al.} \cite{eskildsen} from tunneling data. Note that it has
been suggested  that  for  uniaxial  superconductors  $\tilde{\kappa}$
could     be     approximated     to     $\lambda_{ab}/\lambda_c$
for all field directions \cite{balatskii}  which would lead to $\lambda_c \approx \lambda_{ab} \sim 60$ nm.\

The $\lambda_{ab} (0)$ value determined in this experiment is much
 larger than  the one previously  obtained on single  crystals by
Caplin {\it et al.} \cite{perkins}  but lower than those obtained
on thin  films and polycrystalline  sampes \cite{chen,carrington,buzea}.
However, $\lambda$ is  expected to be very   sensitive  to
interband  scatering and  it has  been shown  \cite{golubov} that
$\lambda_{ab}^{dirty}(0)     \approx      100$     nm     whereas
$\lambda_{ab}^{clean}(0)      \approx       40$      nm      (and
$\lambda_c^{dirty}(0)  \approx 300$  nm and $\lambda_c^{clean}(0)
\approx 40$ nm) suggesting that our single crystals are close to
the clean limit as further confirmed by our  $\Gamma_\lambda$ value
$\approx 1.4$ ($\Gamma_\lambda^{clean}(0) \sim 1$ and
$\Gamma_\lambda^{dirty}(0) \sim 3$). This ratio is also in good agreement
with the one recently deduced from  neutron scattering measurements
\cite{neutron}. \

\begin{figure}
 \includegraphics [width=6cm]{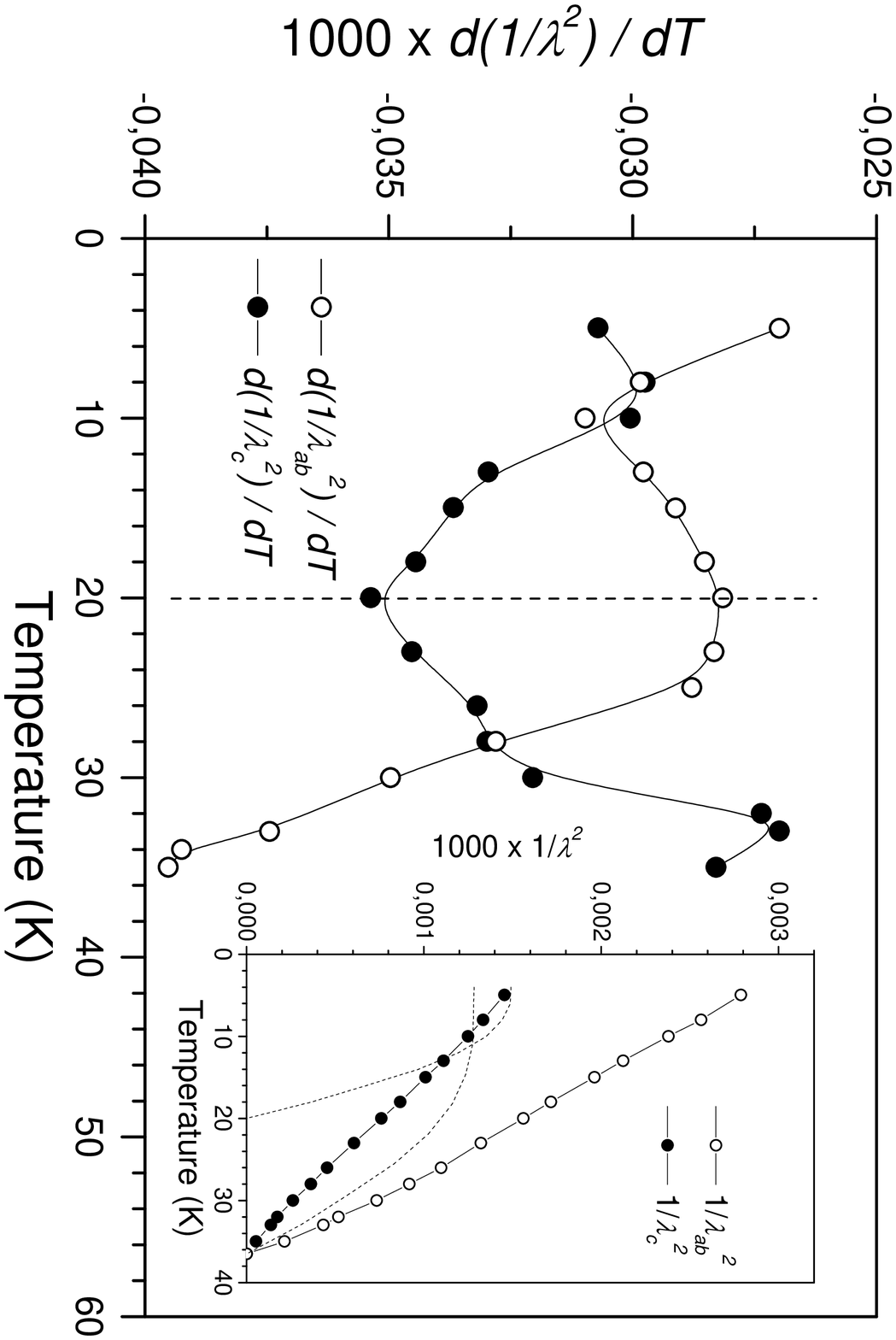}
  \caption{Temperature dependence of $d(1/\lambda^2)/dT$ showing a maximum (resp. minimum) in the $ab-$planes (resp. $c-$direction). In the inset : temperature dependence of $1/\lambda^2$ for the two main crystallographic directions.}
\end{figure}

The temperature dependence of the as-deduced $\Gamma_\lambda$ values
is shown in Fig.3 as solid circles.  $\Gamma_\xi (T)$ (open circles)
has been deduced from our $H_{c2}$ values above $26$ K and  extrapolated
towards lower temperature using the  $H_{c2}(T)$ extrapolation shown
in the inset of Fig.2. As shown, in striking contrast with conventional
one band superconductors
the anisotropy parameters $\Gamma_\lambda (T)$ and $\Gamma_\xi (T)$ are
very different at low temperatures. Moreover, they reveal opposite
temperature dependencies and $\Gamma_\lambda (T)$  increases tending
towards $\Gamma_\xi (T)$ for $T \rightarrow T_c$ where both
anisotropies merges at $\Gamma_\xi (T_c) = \Gamma_\lambda (T_c) \sim 2$.
This unusual behavior is also in striking contrast with the one previously
obtained by Caplin {\it et al}. \cite{perkins} who suggested that
$\Gamma_\xi = \Gamma_\lambda\sim 2$ in the entire temperature range
but is in a very good agreement with  Kogan's \cite{kogan} calculations
of the anisotropy parameters for a weakly-coupled two-bands superconductor.\

At low temperature the anisotropy of the upper critical field is
mainly related to the anisotropy of the Fermi  velocities
{\it  over  the  quasi  2D  $\sigma$  sheet} :
$\Gamma_\xi \approx  \sqrt{<v_{ab}^2>^\sigma/<v_c^2>^\sigma} \sim
6$. $\Gamma_\xi$ then decreases with temperature as the influence
of  the small,  nearly isotropic,  gap increases  due to  thermal
mixing  of the  two gaps \cite{kogan,dahm}.  On the  other hand,  the anisotropy of
$\lambda$ is  related to the  anisotropy of the  Fermi velocities
{\it over the whole Fermi surface} which is expected to be on the
order of $1.1$ in reasonable agreement with our experimental data.
As pointed out by Kogan {\it et al.} \cite{kogan} the two anisotropies
have to merge at $T_c$ as they are then determined by the same "mass tensor".
We  have also  reported in  Fig.3 the  temperature dependence  of
$\Gamma_{H_{c1}}   =   H_{c1    \|c}/H_{c1   \|ab}$.   As   shown
$\Gamma_{H_{c1}}$  and  $\Gamma_\lambda$   present  very  similar
temperature dependencies (and only differ by a numerical factor on
the order  of $1.4$) showing  this dependence does  not depend on
the  choice of  $\tilde{\kappa}$ in  Eq.(1). \

Finally we  discuss the temperature dependence  of the superfluid
density  $\propto  1/\lambda^2$.  The  inset  of  Fig.4  displays
$(\lambda(0)/\lambda(T))^2$ as a function  of temperature for the
two  main crystallographic  directions.  The  dotted lines  schematically   represents   the  temperature dependencies  of  the  superfluid  densities  in  the case of two
independent BCS superconducting $\pi $  (for which $T_c$ would be
on the order of $20$ K)  and $\sigma$ bands. At low temperatures,
the $T-$dependence of  $1/\lambda_c^2$ and $1/\lambda_{ab}^2$ are both
determined by the the band with the smallest gap (i.e. the $\pi -$band)
and $\Gamma_\lambda$ is only very weakly temperature dependent.
As the temperature increases, the influence
of the $\sigma -$ band sets in and an inflexion point is then clearly visible
around $20$K for $1/\lambda_{ab}^2(T)$
(see maximum for $d(1/\lambda_{ab}^2)/dT$ in Fig.4)
followed by a downward curvature up to $T_c $.  As pointed out in \cite{golubov},
this inflexion point is a direct consequence of the "superposition"
of two BCS bands. The sharp kink that would appear for two independent
bands is here smoothed out by interband coupling but still shows up as
a clear maximum in the first derivative close to the "critical temperature"
of the $\pi-$band. A very similar behaviour has been previously
reported by Carrington {\it et al.} \cite{carrington} for the
superfluid density deduced from  microwave measurements. \

It is important  to note that this change in  the curvature at $T
= 20$  K is  also visible  for $1/\lambda_c^2$  (see mimimum  for
$d(1/\lambda_c^2)/dT$   in   Fig.4).   In   this  direction,  the
penetration depth  is mainly determined  by the $\pi-$band  up to
$T_c$ (in  the clean limit)  due to the  very small value  of the
superfluid  density  of  the  $\sigma-$band  (i.e.  strongly  2D
character of the $\sigma-$ band).  Despite the small value of the
corresponding  gap,  superconductivity  remains  induced  in  the
$\pi$ band by interband coupling giving rise to the "tail"
of $1/\lambda_c^2$ at high temperature which shows up as a
lower $d(1/\lambda_c^2)/dT$ value. Those temperature dependencies
are qualitatively in good agreement with the recent calculations
by Golubov {\it et al.}  \cite{golubov}. However, the details
of those dependencies strongly depend on intraband scattering
as well as on the relative contributions of the two bands.

In conclusion,  we have given  direct experimental evidence  that
the  anisotropy  of  MgB$_2$  is  characterized  by two different
parameters ($\Gamma_\lambda$ and  $\Gamma_\xi$) which differs not
only   in  absolute   values  but   also  in   their  temperature
dependencies  in  striking  contrast   to  conventional  type  II
superconductors. The as-deduced temperature dependencies of those
two   anisotropy  parameters   are  in   perfect  agreement  with
theoretical   predictions  for   weakly  coupled    two   bands
superconductor \cite{kogan}.
The presence of two different bands is enforced by the inflexion
point observed around $20$ K in the temperature dependence of
$1/\lambda^2$ in both crystallographic directions.

The  work of P.Sz. was partially  supported  by  the Slovak Science and
Technology     Assistance      Agency     under     contract
No.APVT-51-020102. This work in Korea was supported by the Ministry of Science and
Technology of Korea through the Creative Research Initiative
Program.

\end{multicols}


\begin{references}

\bibitem{nagamatsu}
J.  Nagamatsu et al.,  Nature {\bf 410},63 (2001).

\bibitem{szabo}
P. Szab\'o et al., Phys. Rev. Lett. {\bf 87}, 137005 (2001), F.Giubileo et al. Phys. Rev. Lett. {\bf 87}, 177008 (2001), F.Bouquet et al. Europhys. Lett. {\bf 56}, 856 (2001).

\bibitem{liu}
A. Y. Liu et al., Phys. Rev. Lett. {\bf 87}, 087005 (2001).

\bibitem{kogan}
V.G.Kogan, Phys. Rev. B {\bf 66} 020509(R) (2002); P. Miranovic et al. J. Phys. Soc. Jpn. {\bf 72} 221 (2003).

\bibitem{dahm}
T.Dahm and N.Schopol, Phys. Rev. Lett. {\bf 91} 017001 (2003).
\bibitem{lyard}
L. Lyard et al. Phys. Rev. B. {\bf 66}, 180502(R) (2002).

\bibitem{angst}
S.L. Bud'ko et al. Phys. Rev. B {\bf 64}, 180506 (2001), M.Angst et al. Phys. Rev. Lett. {\bf 88}, 167004 (2002), U.Welp et al. Phys. Rev. B {\bf 67}, 012505 (2003).

\bibitem{neutron}
R.Cubitt et al. Phys. Rev. Lett. {\bf 90}, 157002 (2003).

\bibitem{sungik}
K.H.P.Kim et al. Phys. Rev. B {\bf 65}, 100510(R) (2002).

\bibitem{zeldov}
E.Zeldov et al. Phys. Rev. Lett. {\bf 73}, 1428 (1994).

\bibitem{konczykowski}
M.konczykowski et al. Phys. Rev. B {\bf 43 }, 13707 (1991).

\bibitem{buzea}
For a review see : C. Buzea et al. Supercond. Sci. Technol. {\bf 14} R115 (2001)

\bibitem{askerzade}
I.N.Askerzade et al. Supercond. Sci. Techno. {\bf 15} L17 (2002)

\bibitem{abrikosov}
A.A.Abrikosov, in book {\it Fundamentals of the theory of metals}, North-Holland, Amsterdam (1988).

\bibitem{samuely}
P.Samuely et al. Physica C {\bf 385}, 154 (2003).

\bibitem{bouquet}
F.Bouquet et al. Phys. Rev. Lett. {\bf 89}, 257001 (2002).

\bibitem{eskildsen}
M.R.Eskildsen et al. Phys. Rev. Lett. {\bf 89} 187003 (2002).

\bibitem{balatskii}
A.B. Balatskii et al. Sov. Phys. JETP {\bf 63} 866 (1986).

\bibitem{perkins}
A.D.Caplin et al. Supercond. Sci. Technol. {\bf 16} 173 (2003).

\bibitem{chen}
X.H.Chen et al. Phys. Rev. B {\bf 64}, 172501 (2001), C.Niedermayer et al. Phys. Rev. B {\bf 65} 094512 (2002).

\bibitem{carrington}
F.Manzano et al. Phys. Rev. Lett. {\bf 88} 047002 (2002), A.Carrington and F.Manzano, Physica C {\bf 385}, 205 (2003).

\bibitem{golubov}
A.A. Golubov at al. Phys. Rev. B. {\bf 66}, 054524 (2002).

\end{references}
\end{document}